\documentclass{article}

\usepackage{arxiv}

\usepackage[utf8]{inputenc} 
\usepackage[T1]{fontenc}    
\usepackage{hyperref}       
\usepackage{url}            
\usepackage{booktabs}       
\usepackage{amsfonts}       
\usepackage{nicefrac}       
\usepackage{microtype}      
\usepackage{lipsum}
\usepackage{graphicx}

\title{MakeSense: An IoT Testbed for Social Research of Indoor Activities}

\author{
  Jie Jiang\\
  University of Surrey\\
  Guildford, GU2 7XH \\
  \texttt{jie.jiang@surrey.ac.uk} \\
   \And
  Riccardo Pozza \\
  University of Surrey\\
  Guildford, GU2 7XH \\
  \texttt{r.pozza@surrey.ac.uk} \\
   \AND
  Nigel Gilbert \\
  University of Surrey\\
  Guildford, GU2 7XH \\
  \texttt{n.gilbert@surrey.ac.uk} \\
   \And
  Klaus Moessner \\
  University of Surrey\\
  Guildford, GU2 7XH \\
  \texttt{k.moessner@surrey.ac.uk} \\
}

\begin{document}
\maketitle

\begin{abstract}
There has been increasing interest in deploying IoT devices to study human behaviour in locations such as homes and offices. Such devices can be deployed in a laboratory or `in the wild' in natural environments. The latter allows one to collect behavioural data that is not contaminated by the artificiality of a laboratory experiment. Using IoT devices in ordinary environments also brings the benefits of reduced cost, as compared with lab experiments, and less disturbance to the participants' daily routines which in turn helps with recruiting them into the research.  However, in this case, it is essential to have an IoT infrastructure that can be easily and swiftly installed and from which real-time data can be securely and straightforwardly collected. In this paper, we present MakeSense, an IoT testbed that enables real-world experimentation for large scale social research on indoor activities through real-time monitoring and/or situation-aware applications. The testbed features quick setup, flexibility in deployment, the integration of a range of IoT devices, resilience, and scalability. We also present two case studies to demonstrate the use of the testbed, one in homes and one in offices. 
\end{abstract}


\section{Introduction}
Internet of Things (IoT) is a paradigm encompassing various application areas, ranging from smart homes and buildings, intelligent transportation systems and smart cities to industrial automation, smart healthcare and smart grids \cite{Al-Fuqaha:2015}. In the domain of smart homes and buildings, IoT mainly advocates the need for distributing multi-modal sensing and actuation capabilities into our living environments (e.g. homes, offices), in order to make the objects we interact with everyday smarter and thus improve our lives. In such scenarios, being aware of the contexts or surrounding environments \cite{Perera:2014} and of the daily activities \cite{Wang:2018} carried out by the occupants not only help in designing smart environments and intelligent applications but also allow social researchers to study in a larger scale, for example, household or office practices in real world settings. 

Social researchers take a great interest in understanding real-life practices, e.g., family dynamics and child-rearing (e.g., \cite{Ruppanner:2015,Atkinson:2016}), practices around meals \cite{Gattshall:2008}, sleep \cite{Williams:2015}, assisted living arrangements and mobile health solutions (e.g., \cite{Lopez-Gomez:2015,Mort:2013}), homeworking \cite{Sullivan:2016} and energy-related practices \cite{Pierce:2010}. In office environments, monitoring the indoor environment quality and the productivity of the occupants is another area of research \cite{AlHorr:2016} where matters such as, personal thermal comfort, levels of ambient lighting, noise and air quality are some of the main properties requiring monitoring and control to ensure the well being of office occupants.

Existing social research methods are both qualitative and quantitative, and often some combination of the two are used for pragmatic and constructivist purposes \cite{Shannon-Baker:2016}.
With the advance of IoT, sensor-generated data are becoming widely available and the use of such data for social studies of indoor activities has thrived in recent years with applications in areas such as smart homes and offices, assisted living, etc. For example, Williams et al. \cite{Williams:2015} discuss the use of accelerometers to study people's sleep patterns. Amft and Tr\"{o}ster \cite{Amft:2008} study people's dietary behaviour by using inertial sensors to recognise movements, a sensor collar for recognising swallowing and an ear microphone for recognising chewing. Wang et al. \cite{Wang:2014} help in detecting elderly accidental falls by employing accelerometers and cardiotachometers. 
Microsoft used a multimodal system \cite{Nuria:2005} for office activity recognition, recognizing user state from cameras, audio sensing and computer keyboard and mouse interactions. 

In the literature, there are in general two types of IoT infrastructure that have been used by social researchers to supplement survey, questionnaire, interview, time use diary, etc.   
The first type of IoT infrastructure is embedded in the concept of living labs which are an initial attempt to carry out social studies in realistic environments. The Philips HomeLab \cite{deRuyter:2005} and the PlaceLab \cite{Intille:2006} are two examples of such living labs that provide a living space where the participants agree to stay there for a temporary period of time and have their activities and social interactions being monitored. These living labs, though enable to study constructed social environments, require people to temporarily leave their original real world settings (e.g. homes and offices) and behave as if they were in the real world habitats. A drawback of the living lab method is that the experimental settings might not be able to account for the diversity of the environments and behaviours existing in the real world. Moreover, this method cannot scale due to the cost of the infrastructure and it is also difficult to recruit participants if the experiments require a long period of continuous participation, e.g. several months.  
The second type of IoT infrastructures are the smart home testbeds developed in for example the CASAS project \cite{Cook:2013}, CARE project \cite{Krose:2008} and Sphere project \cite{Zhu:2015}. These projects appeared to be the early efforts to design sensing and actuating infrastructures that can be deployed into real homes for monitoring people's daily living for research purposes such as health monitoring.
Nevertheless, these testbeds focus more on the analytical aspects of sensor-generated data rather than providing an easy-to-use IoT infrastructure for social researchers to build their own applications for studying different aspects of indoor activities. Moreover, the design of these testbeds is mostly tailored to showing the viability of using sensor-generated data in specific research domains and thus provide limited support for sensor integration, privacy control, connectivity, data visualisation and etc.

For social studies of indoor activities to be carried out smoothly, it is essential that an IoT testbed meets the following requirements: 
\begin{itemize}
\item The IoT testbed should be able to tailored to diverse experiment settings.
\item The IoT devices should be non-intrusive, cost-efficient and easy to deploy.
\item The IoT devices should not interfere with or disturb the participants' daily routines.
\item The privacy of the participants should be guarded at a high standard during and after the data collection.
\item The experiment coordinators should be able to monitor the data collection. 
\end{itemize}

To these ends, in this paper we introduce MakeSense, an IoT testbed that aims to facilitate the experimentation of social research in real world indoor environments such as homes and offices. The features of MakeSense can be summarized as follows:
\begin{itemize}
\item The testbed supports extensible APIs for integrating additional self-developed and commercial sensors.
\item The non-intrusive design of the testbed facilitates user acceptability by excluding image/video based IoT devices and minimises the disturbance to the participants' daily routines.
\item The IoT devices comprises commodity sensor modules to minimise the material cost.
\item The testbed provides pre-built Docker images and scripts \footnote{https://github.com/jiejiang-jojo/make-sense} to automate the setup and deployment process.
\item The on-board WiFi/Bluetooth enable quick and flexible deployment of the devices in complex environments.
\item End-to-end data encryption and pseudonymisation are employed to preserve the security and privacy of the participants.
\item The testbed supports both real-time data transfer and local storage to enable real-time monitoring and ensure the quality of data collection.
\end{itemize}

The rest of the paper is organised as follows. In Section \ref{sec:rel}, we discuss the related work. In Section \ref{sec:mak}, we introduce the IoT testbed, namely MakeSense, and illustrate its sensor network, mechanism of data collection, data management and monitoring, data visualisation and analysis. In Section \ref{sec:showcases}, we present two use cases showing how MakeSense was used to assist two research projects regarding home and office monitoring. Thereafter, in Section \ref{sec:dis}, we share some lessons learned from the two projects. Finally, we conclude our work with some possible extensions in Section \ref{sec:con}.

\section{Related Work}
\label{sec:rel}

Recent technological advances have resulted in cost reduction and increased
availability of hardware needed for conducting real-life experiments at large-scales, which drives the need for IoT testbeds to enable faster experimentation with various sizes, hardware, topologies and degrees of flexibility \cite{TONNEAU2015}.   
In the literature, a vast number of testbeds were developed to facilitate wireless sensor network experimentation. For example, one of the most comprehensive IoT testbeds SmartSantander \cite{SmartSantander2014} provides an experimental research facility that could be used to test smart city applications and services. It offers thousands of fixed and mobile sensors (environment, parking, transportation, etc.). Some other examples of testbeds targeting at smart city applications include SoundCity \cite{SoundCity2016}, FINE \cite{FederationIoT2018, FINE:2019}, etc. 
Though these testbeds are powerful in aspects such as scale, heterogeneity, mobility, etc., they are often complex to set up and requires a lot of technological training. 
Another group of testbeds similar to ours focus on the experimentation of smart building applications. For example, the SmartCampus testbed \cite{SmartCampus2013} provides a facility for IoT experimentation using a variety of sensor devices deployed within buildings that could measure temperature, humidity, illuminance, noise, air quality, occupancy and energy consumption. Similarly, the KETI testbed aims to collect sensing data from offices and parking lots by deploying sensors that could measure indoor climate, energy consumption of office utilities, people's presence in offices and parking lot status. These two testbeds are similar to ours in terms of the provided IoT devices and domains of application. However, neither of them offers an end-to-end solution from data collection to data visualisation with automated setup and deployment support.  

Living labs were another popular way of conducting research of indoor activities in realistic settings, and began to flourish in the last decade. An early establishment is the Philiphs HomeLab \cite{deRuyter:2005} which provides behavioral researchers an instrument for studying human behavior in a home setting with cameras and microphones installed throughout the home. Another example is the MIT PlaceLab \cite{Intille:2006} which offers a facility for the study of ubiquitous technologies in home settings via a network of sensors that could capture a complete record of audio-visual activity, including information about objects manipulated, environmental conditions, and use of appliances. 
Some other examples of such living labs include the Gator Tech Smart House \cite{Helal:2005} from the University of Florida, the MIT Human Speechome Project \cite{Roy_Home2006}, the Assisted Living Laboratory \cite{Kleinberger:2007} at the Fraunhofer IESE in Kaiserslautern, the Georgia Tech Aware Home \cite{Kientz:2008}, the TigerPlace \cite{Skubic:2009} in Columbia, Missouri and more recently the Halmstad Intelligent Home \cite{Halmstad2016} at Halmstad University and the UJAmI \cite{Espinilla:2018} at the University of Ja\'en, Spain. Most of these labs however have the drawback of using intrusive sensors such as cameras. Even though in some cases the cameras were used to capture just silhouettes, privacy issues still stand since the data can be used to extrapolate contextual information about occupants. 
Another drawback of the living lab approach is that the experimental settings might not be able to account for the diversity of the environments and behaviours existing in the real world. Moreover, this method cannot scale due to the cost of the infrastructure and it is also difficult to recruit participants if the experiments require a long period of continuous participation, e.g. several months.

There is another relevant group of works focusing on home monitoring and activity recognition. For example, the Ambient Assisted Living Context Aware Residence for Elderly (CARE) \cite{Krose:2008} project proposed a non-intrusive sensing platform by leveraging proximity and switch like sensors to recognise activities from the elderly living alone at home. The Washington State University CASAS \cite{Cook:2013} project tried to move away from living laboratories with easy-to-install battery operated devices deployed in homes such as temperature sensor, door sensor, Infrared
motion/light sensor. In the SPHERE project \cite{Zhu:2015}, a platform based on wearables, video cameras and environment sensors for healthcare in residential environments, was deployed in 100 households in Bristol, UK. Targeting on general-purpose sensing, the Synthetic Sensors \cite{Laput2017} developed by the researchers from Carnegie Mellon University integrate a set of non-intrusive sensor modules and proposed to apply machine learning methods to recognise meaningful events at different abstraction levels from raw sensor data. 
Though these works all provide a solution for home monitoring with different sets of sensors, their focus is either on the sensing technology itself and/or on researching methods for activity recognition rather than developing an IoT infrastructure that could be easily adjusted and deployed for social research of indoor activities in different settings such as homes or offices.

\section{MakeSense}
\label{sec:mak}
MakeSense has been designed to provide an infrastructure that enables real-world experimentation for large scale social research of indoor activities through real-time monitoring and/or situation-aware applications. The use of cloud infrastructure for data monitoring and storage, as well as for data analysis and data visualization introduces benefits for scalability, ease of extension, reliability and manageability. 
The design of the self-manufactured IoTEgg device, developed for MakeSense, leverages commodity ambient sensors and moves away from intrusive IoT devices, for example, cameras. Furthermore, due to its multi-radio setup with Bluetooth Low Energy (BLE) and WiFi, MakeSense allows ad-hoc self-manufactured or commercial sensors to be added and to be interfaced with the IoTEggs acting as relays. This can be achieved easily by exploiting MakeSense's standard APIs and providing access to sensors and actuators via its IoT protocol, the Open Mobile Alliance (OMA) lightweight M2M (LWM2M) protocol \cite{Lwm2m:2018}, which allows to interact and remotely manage in a secure and efficient way resource constrained IoT devices.

\subsection{Testbed Architecture}
\begin{figure}[!h]
\centering
  \includegraphics[width=\textwidth]{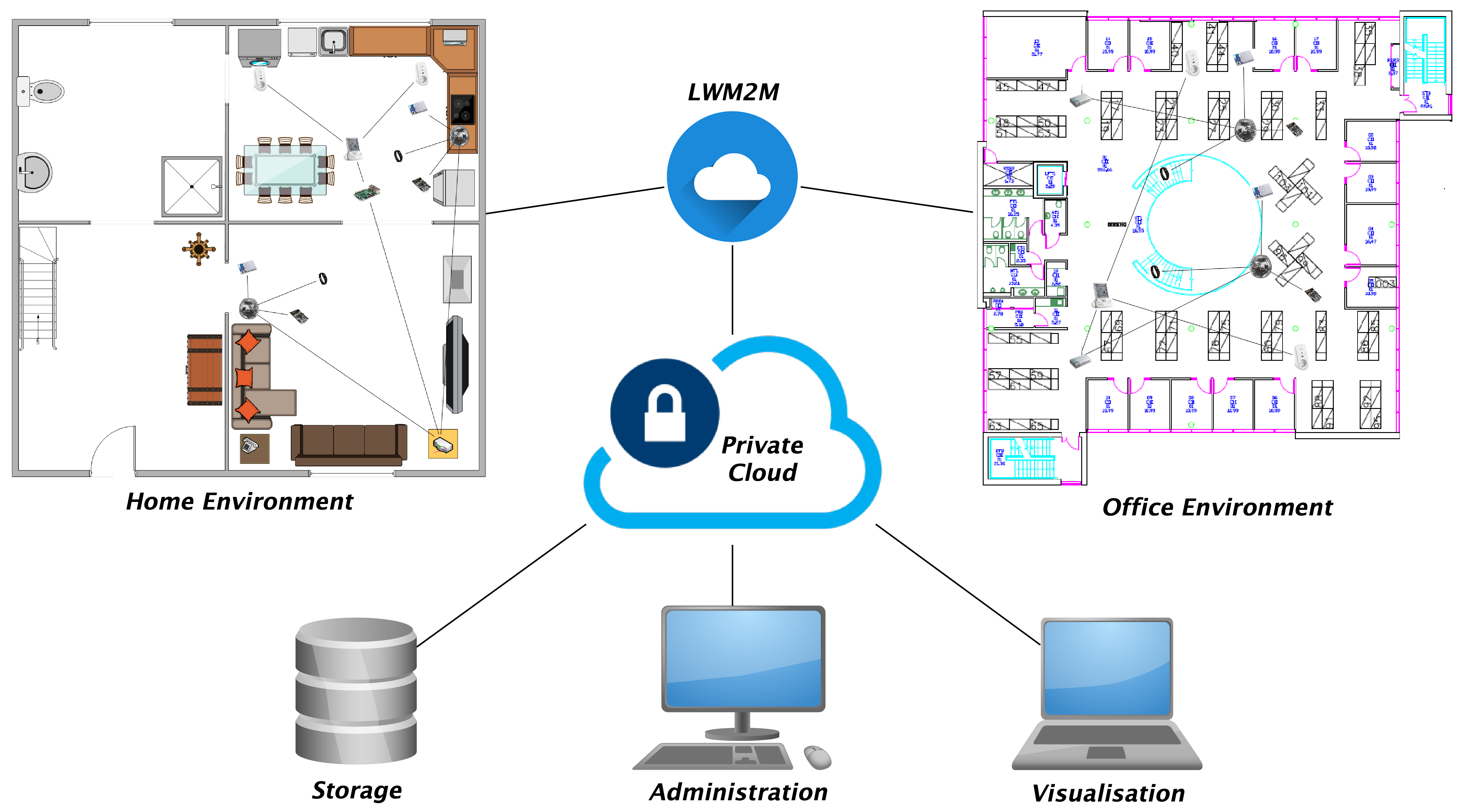}
  \caption{MakeSense Network Layout.}
  \label{fig:nw-layout}
\end{figure}

MakeSense adopts a client-server like paradigm. All the sensing devices deployed in indoor environments such as homes or offices, as shown in the top row of Figure \ref{fig:nw-layout}, serve as clients that collect physical environment data which are sent directly to the server backends via LWM2M through Broadband Internet. The server backends can be hosted on a private cloud or an on-premise dedicated server with enough capacity (as shown in the centre of Figure \ref{fig:nw-layout}). Besides receiving data from sensors and storing them on persistent storage, the server backends also provide support for administration and visualisation through web applications. In the rest of this section, we will elaborate on the technical design choices of MakeSense.

MakeSense features both commercial off-the-shelf (COTS) IoT devices, i.e. energy monitors and BLE wristbands, and self-manufactured IoT devices, i.e., the IoTEggs (see Sections \ref{sec:selfsensors} and \ref{sec:cotssensors} for more details). These IoT devices can be directly connected to WiFi access points or benefit from the capability of the IoTEggs' WiFi modules to operate in both station and access point modes, as well as from their on-board BLE module. Both static and mobile (e.g. wristbands) COTS and self-manufactured BLE and WiFi based sensors/actuators can thus be utilized by the IoTEggs, augmenting their on-board default set of sensors/actuators and enabling opportunistic sensing interactions.
Data captured by the IoT devices are collected from houses or buildings relying on readily available WiFi networks and Internet access, thus benefiting from higher reliability, lower interference and higher throughput as compared with other radio technologies that would need to be deployed, i.e. other lower power radios.
Concerning power supply, nowadays, most houses and buildings have sufficiently distributed power sockets, which makes the powering of IoT devices such as the IoTEggs easy. Moreover, to reach locations not easily accessible or subject to mobility, BLE based sensors may also be used, but they have to be charged periodically. 

MakeSense supports data transmission and device management over LWM2M which is used together with the Constrained Application Protocol (CoAP) \cite{Coap:2018}, an application layer protocol designed for resource constrained IoT devices. 
By using WiFi and CoAP via UDP protocol, high temporal resolution measurements can be collected, limited only by the sampling requirements of the physical sensors. Furthermore, MakeSense implements security at the communication level with Datagram Transport Layer Security (DTLS) and pre-shared key authentication, and all the data are encrypted with Advanced Encryption Standard (AES) 128 bit before transmission.

The server side is deployed on a private cloud that supports persistent storage of sensor-generated data via databases deployed on virtual machines (VMs). The infrastructure is based on OpenStack, with storage distributed identically between disk/volume based (Cinder) and document based object (Swift) storage services. Moreover, front end interfaces for monitoring, administration and visualisation are deployed inside VMs that can be easily accessed anywhere through the Internet via web browsers. The use of a private based cloud solution was chosen because it enables scalable and efficient handling of large scale experiments, thus reaching a large number of users.

MakeSense introduces a reliable storage solution by providing data backup and replication. On the remote cloud infrastructure, data storage serves two purposes, i.e., data post-processing and real-time data visualisation/analysis. In the former case, the open-source databases such as PostgreSQL \cite{PostgreSQL:2018} and MongoDB \cite{MongoDB:2018} are used. In the latter case, the open-source search engine Elasticsearch \cite{ElasticSearch:2018} is used together with its dedicated visualisation platform Kibana \cite{Kibana:2018}. MakeSense opts out the use of Logstash \cite{Logstash:2018}, a data loading service commonly used together with Elasticsearch, and instead employs a light-weight replacement based on Python. Users may adjust the choice of these database and visualisation/analysis modules according to their needs. 

\begin{figure}[!h]
\centering
  \includegraphics[width=0.8\textwidth]{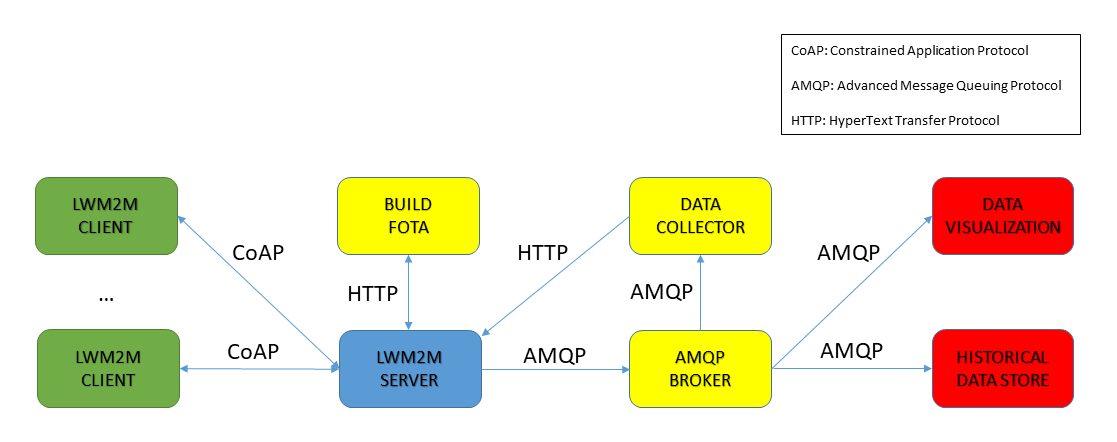}
  \caption{MakeSense Software Architecture.}
  \label{fig:sw-arch}
\end{figure}

Figure \ref{fig:sw-arch} shows the software architecture of MakeSense. 
Since LWM2M adopts a client-server architecture, we developed a LWM2M client running on the IoTEgg, interfacing with the embedded sensors and actuators, and communicating via CoAP with a LWM2M server. In order to provide a model for interoperability between devices and  applications/services deployed on the cloud side, the sensors and actuators were developed as IPSO Smart Objects \cite{IPSO:2016}. 

By leveraging LWM2M and IPSO smart objects, M2M applications are deployed on the cloud and operated using standard APIs. LWM2M clients maintain their registration and thus their current availability as well as the list of their objects and resources (i.e. sensors and actuators) via a LWM2M server registry, accessible via HTTP requests.
For MakeSense, as shown in Figure \ref{fig:sw-arch}, we designed two M2M applications interfacing with the LWM2M Server via HTTP: one for building images and downloading firmware over-the-air (FOTA) to the IoTEggs and the other for handling data collection from the IoTEggs, configuring the notification interval and initiating requests for data collection (if the device or sensor is not blacklisted) via the standard LWM2M Observation/Notification interface upon device registration.

The LWM2M server publishes via Advanced Message Queuing Protocol (AMQP) to a RabbitMQ message queue broker both control messages (e.g. LWM2M client registrations) on one queue exchange and live sensor measurements on another queue exchange. 
The use of a message queue is for asynchronous communication, increased reliability and scalability as well as decoupling between the data collection and data storage/visualisation applications.
In addition to the M2M application for data collection which subscribes to the control messages queue exchange, we designed another two applications that instantiate multiple workers which collect sensor measurements from the live data message queue exchange and store them into historical data stores (e.g. PostgreSQL and MongoDB) and into ElasticSearch for subsequent visualisation and analysis with Kibana.

\subsection{Sensor Network}
MakeSense supports a range of connectivity options enabling virtually any sensor capable of interfacing with BLE and WiFi to be used in conjunction with the IoTEgg acting as a gateway. Given both the non-intrusive design of the IoTEgg and the secured communication and storage, the security and privacy of the participants of the study is preserved. Furthermore, due to its scalable design in terms of both sensor variety and ease of extension in the cloud, MakeSense enables research experiments to be carried out in a wide range of real residential as well as commercial indoor environments.

\subsubsection{Self-Manufactured Sensor Suites} 
\label{sec:selfsensors}
\begin{figure}[!h]
\centering
  \includegraphics[width=0.25\textwidth]{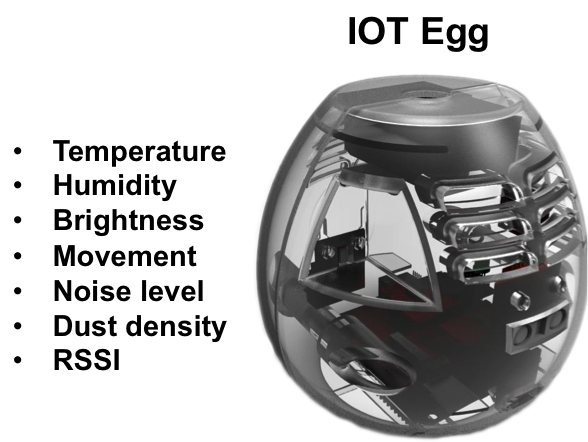}
  \caption{IoTEgg consisting of a suite of sensor modules}
  \label{fig:iotegg}
\end{figure}

In the market, there is a vast number of electronic modules that can be used to build sensor suites for different measurements. One example of such development is the IoTEgg designed and developed at the 5G Innovation Centre (5GIC) at the University of Surrey. The IoTEgg is encapsulated in a box, as shown in Figure \ref{fig:iotegg} (a), and is coordinated by a Seeeduino Arch-Pro development board \cite{Seeeduino:2018}, based on an NXP LPC1768 microcontroller unit (MCU). A temperature and humidity sensor board HTU21D \cite{TempHumi:2018} is managed via an I2C interface and sampled periodically by the client application deployed on the ARM core. An Avago ADPS-9960 light sensor \cite{Light:2018}, also managed via an I2C interface, is used to sample ambient light measured in $\frac{\mu W}{cm^2}$. The GP2Y0A60SZ ranging sensor from Sharp \cite{Range:2018} included in the design, is an analog sensor with a wide detection range of 10 cm to 150 cm, which is sampled via a 12 bit ADC and converted through the manufacturer's calibration table. Finally, a MEMS microphone breakout board INMP401 \cite{Microphone:2018} is used to sample noise levels in the environment on another ADC channel and the values are converted to sound-pressure-level decibels (dB SPL). The sampling rate of the IoTEgg can be customised for different use cases via a configuration file. 

The IoTEgg has been designed, prototyped and manufactured internally in our centre as a means for both capturing environmental changes in physical measures and acting as a hub for relaying data from and to additional wireless sensors. 
The IoTEgg has been equipped with a Bluetooth Low Energy (Bluegiga BLE112) module that allows additional COTS or self-manufactured sensor suites to be added and connected to the IoTEgg, so to enable further data collection. Furthermore, the WiFi module (ESP8266) allows both operation either as a WiFi access point or as a WiFi Station, or both together at the same time, thus enabling also additional WiFi based sensors to be connected to the IoTEgg for relaying data. 

This setup has the clear advantages of not requiring much effort for installation, the infrastructure can be deployed in minutes, and of allowing practical extensibility with COTS or ad-hoc battery operated devices connected to the IoTEgg working as a gateway, as well as allowing local storage in the IoTEgg.
For example, static wireless devices can be attached to fixed real world objects (e.g. motion sensors to drawers, doors) or mobile real world objects (e.g. wristbands and smartphones to people), thus capturing broader dynamics.

One may assemble a combination of different electronic modules to build sensor suites according to their needs. The advantages of self-manufactured sensor suites are (1) a wide selection of modules are available to choose from, (2) the data collection of each module can be adjusted for specific uses, (3) the cost of the sensors can be controlled by choosing products of varying degrees of sensitivity. On the other hand, there are several disadvantages of self-manufacturing sensor suites: (1) manufacturing requires solid engineering background to be able to design low-cost and high quality sensors, (2) there is usually a fairly long development and test cycle, (3) the cost of sensors might be higher than for mass production, as for larger volumes, costs can be reduced.

\subsubsection{Commercial Off-the-shelf Sensor Suites}
\label{sec:cotssensors}
\begin{figure}[!h]
\centering
  \includegraphics[width=0.5\textwidth]{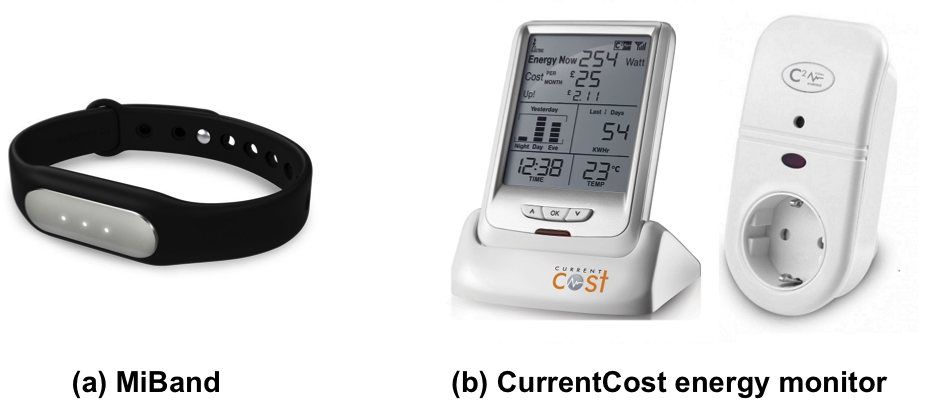}
  \caption{Examples of Commercial Sensor Suites}
  \label{fig:commercialsensors}
\end{figure}

When building one's own sensor suite is not a choice, there is also a wide selection of commercial products ranging from elementary ones like temperature and humidity or accelerometer sensor suites (e.g. Ti Sensor Tags \cite{SensorTag:2018}) to more sophisticated ones like energy monitors, pet tags, smart locks, wearable trackers. Figure \ref{fig:commercialsensors} shows two examples of commercial products that have been used in a number of studies.

Smart wristbands have been widely used by researchers in health care applications, such as activity recognition \cite{Anazco:2018}, fall detection \cite{Khojasteh:2018} and gait analysis for Parkinson's disease \cite{Mazilu:2015}.
Figure \ref{fig:commercialsensors} (a) shows a BLE wristband called MiBand \cite{Miband:2018} which is mainly used by people to track their levels of exercise, e.g., how many steps a person has achieved per day which can be converted to calories. 

However, as for many other commercial products, there are no open APIs available to download data for further analysis, which makes it more challenging for researchers to harvest the data for research purposes. In our case, instead of using the wristband for its original purpose, we exploited the BLE module built in both the wristband and the IoTEgg to capture the received signal strength indication (RSSI) distribution over time, so to have a rough localization method. This allowed us to estimate how close the wristband is to the IoTEggs placed in different locations of a building based on the RSSI measured by each IoTEgg. The combined use of the IoTEgg and the wristband is an example of how other sensors might be integrated to the sensing capabilities of the IoTEgg.   

Another example of commercial sensor suite is the electricity monitoring kit from CurrentCost \cite{CurrentCost:2018} as shown in Figure \ref{fig:commercialsensors} (b), which has been used in several studies such as \cite{Kelly:2015} and \cite{Zhu:2015}. It features a current transformer clamp, a transmitter and a number of individual appliance monitors (IAMs) to measure the energy consumption of the whole house as well as the individual appliances, and a monitor hub to collect and display the sensor data. The measurements are in units of Watts and the maximum sampling rate is every 6 seconds. The users can direct the data stream from the electricity monitors to their own devices for post processing via the standard data port and specification provided by CurrentCost.

The MiBand and the CurrentCost energy monitor presented in this paper are only two examples of integrating commercial sensors with MakeSense. There are a broad range of commercial products that could be explored. For example, Bluetooth enabled activity trackers like MiBand are widely available in the market such as Fitbit \cite{Fitbit:2019} and Garmin \cite{Garmin:2019}. Energy monitoring products are also gaining popularity for both home and business applications. Among the many energy monitors in the market, EmonPi \cite{emonPi:2019} provides an open-platform energy monitoring solution. It supports data access via the open-source web-app Emoncms that could be run locally on the EmonPi itself or users' own server. Another energy monitor product that provides data access is Neurio \cite{Neurio:2019}. Users could direct the data to their own Cloud infrastructure via Neurio Software and Cloud API. With air pollution being a common problem around the globe, air quality monitors are also becoming popular. An example of such a product is Kaiterra \cite{Kaiterra:2019} which measures PM2.5 and TVOC. It supports data access via download from the Kaiterra App. 

With the advance of IoT, more commercial sensing products are becoming available in the market. The main advantages of using commercial products are that (1) typically less development work is required, (2) the sensors are normally more robust, being thoroughly tested, (3) they sometimes rely on ready-to-use platforms for data monitoring. On the other hand, there are several disadvantages: (1) rigid and limited control for data collection, (2) the combination of sensor modules can not be easily changed or extended to support new sensors, (3) cost might be higher for deployments in large volumes, (4) the products may run out of stock or even be out of production.

\subsection{Data Collection}

Nowadays most houses and buildings are covered by WiFi access points which can be used without the need to install dedicated gateways or alternative solutions. As such, MakeSense aims to facilitate data collection through easy-to-setup, fast and reliable WiFi networks. Furthermore, the ease of adding other types of Bluetooth or WiFi enabled sensors and actuators is supported by the IoTEgg, which can be used to relay measurements or control messages back and forth from the cloud infrastructure. 

In order for MakeSense to provide most of the desirable features of an IoT Testbed (i.e. remote management, reprogramming, customisable data collection, live query for location and connectivity, security, etc.), we leverage LWM2M over CoAP and encryption with DTLS and pre-shared keys authentication. 
Moreover, it is important that all the data are pseudonymised from the time they are generated, making it difficult to trace the participants. MakeSense implements data pseudonymisation by assigning random identifiers for each participant and the links between the identifiers and the participants are stored separately.

Since the IoTEgg is an IoT device with limited resources, we ported the Eclipse Wakaama \cite{Wakaama:2018} C library (designed to be build into linux targets) on our resource constrained target (i.e. the Seeeduino Arch-Pro), which runs no operating system and just some libraries as hardware abstraction layer to interface with peripherals. The library adopts a CoAP C implementation which is an adaptation of the Contiki's Erbium CoAP library. 

A full fledged LWM2M client has been developed, featuring the following LWM2M Objects (with standard object numbers): LwM2M Security, LwM2M Server, Access Control, Device, Connectivity Monitoring, Firmware, Location and Connectivity Statistics. For each of these objects' resources, API and drivers have been implemented in C++ and used by Wakaama to be provided as callable endpoints from any LWM2M server. This allows for configuring the device remotely and providing features such as setting security mode (in our case AES128 was used, provided by the tinyDTLS C library), default device lifetime, notification period, static GPS coordinates, time synchronization, factory reset and remote reboot. Furthermore, network and connectivity parameters and statistics can be inquired from remote such as WiFi RSSI, IP addresses and configuration. Finally, APIs for firmware-over-the-air updates have been customised to allow reliable download of new software updates for the LWM2M client via an ad-hoc secondary boot loader which exploits self-programming via an external flash memory in the IoTEgg.

\begin{figure}[!h]
\centering
  \includegraphics[width=0.8\textwidth]{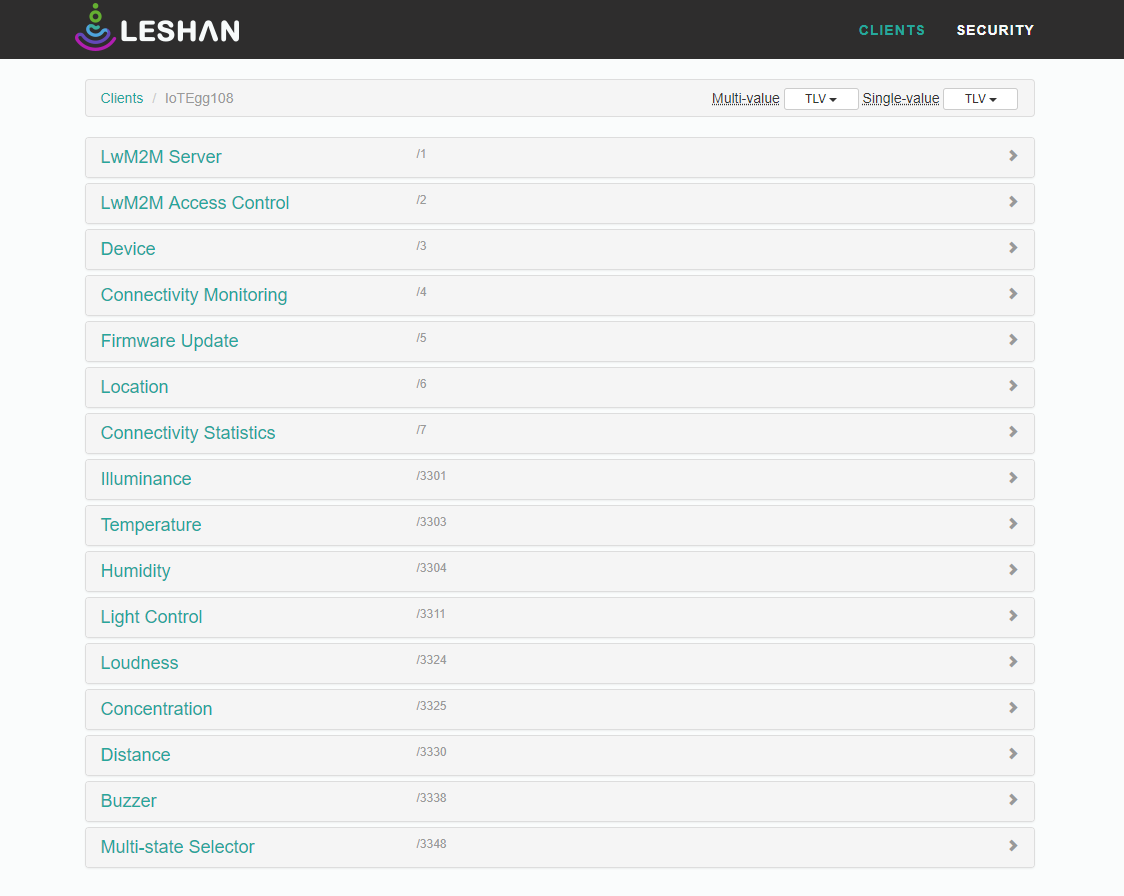}
  \caption{OMA LWM2M Server Objects in MakeSense.}
  \label{fig:lwm2m_server}
\end{figure}

On the server side, as shown in Figure \ref{fig:lwm2m_server}, the LWM2M server (running the Eclipse Leshan \cite{Leshan:2018} Java library) is able to access not only the standard LWM2M objects, but also the additional IPSO Smart Objects (with standard object numbers), representing ambient light luminosity (3301) in $\mu W/{cm}^2$, temperature (3303) in celsius degrees ($^{\circ}C$), humidity (3304) in percentage ($\%$), RGB led control (3311), loudness (3324) in sound-pressure level decibels ($dB$ $SPL$), dust concentration (3325) in $mg/{mm}^3$, proximity distance (3330) in $cm$, buzzer control (3338) and gesture recognition for which we adopted the multi-state detector (3348) where each integer from 0 to 6 represents directions for the gestures.

Some of the advantages of this design are that (1) users of MakeSense could easily design M2M applications themselves to collect data in various data formats and time intervals or act upon the IoTEgg actuators in real-time, (2) updates can be sent over-the-air with minimal disruption, (3) monitoring can also be done remotely to identify issues, (4) the LwM2M standard allows for ease of extension and promotes interoperability due to the use of standard IPSO Smart Objects (see LWM2M specifications for more details), and (5) both data pseudonymisation and encryption are implemented to provide efficient data security.

\subsection{Data Management and Monitoring}
All of the cloud data management is handled via a RabbitMQ message queue broker, deployed on a VM and secured with access control and authentication mechanisms. The broker has three main subscribers listening for data: the data collector service, the data visualisation service and the historical data store service. The data published to the broker comes from two main sources: the control messages (e.g. LWM2M registration updates) and the live data stream from the sensors.

The data collector M2M application, running on a VM as a reliable and always-on  linux service, has been developed in Python. This application listens for the LWM2M client registrations to the LWM2M server via the RabbitMQ, where the server publishes registration messages. The application is thus aware of when devices come online and when data collection from the IPSO Smart Objects can be started by sending requests to the LWM2M server via the HTTP interface. Furthermore, this application helps the IoTEggs synchronise the time with that of the VM.

The data visualisation application is also running on VMs where ElasticSearch and Kibana have been deployed. This has been designed as a reliable and always-on  service in Python, running as multiple instances in parallel. Live sensor data coming from the RabbitMQ is stored in disk drives mounted as OpenStack volumes via the cinder service. Concerning the historical data store application, a similar  service was designed, with storage based on databases such as MongoDB and PostgreSQL.

A M2M application for build generation and firmware over-the-air download has also been designed. The GNU ARM Embedded toolchain has been customised to generate both the secondary boot loader (SBL) and the main LWM2M client application allocated in two separate memory intervals in the MCU internal flash memory. The M2M application thus generates the build image, uploads it to the reliable Openstack Swift object storage service, injects a magic number (i.e. a unique identifier that will be to be cross-verified when the build image is downloaded) and initiates a reboot on the corresponding LWM2M client/s to prepare for a reliable software update via the SBL at the next reboot.

MakeSense integrates the data storage module (e.g. PostgreSQL) and the data visualisation module (Elasticsearch and Kibana) by providing pre-built Docker \cite{Docker:2018} images and scripts that automate the server deployment process \footnote{https://github.com/jiejiang-jojo/make-sense}. However, to be able to use PostgreSQL to make data queries or use Kibana to create data visualisations, one needs to be familiar with the basics of SQL and Kibana visualisations.

Some of the advantages of this architecture are that (1) users of MakeSense can easily subscribe to the broker and add their workers to implement live data processing and analysis, (2) MakeSense provides both historical data storage and short term storage for live visualisation, (3) remote control of the IoT devices makes it easy for software updates and thus facilities infrastructure maintenance.

\subsection{Data Visualisation and Analysis}
\begin{figure}[!h]
\centering
  \includegraphics[width=\textwidth]{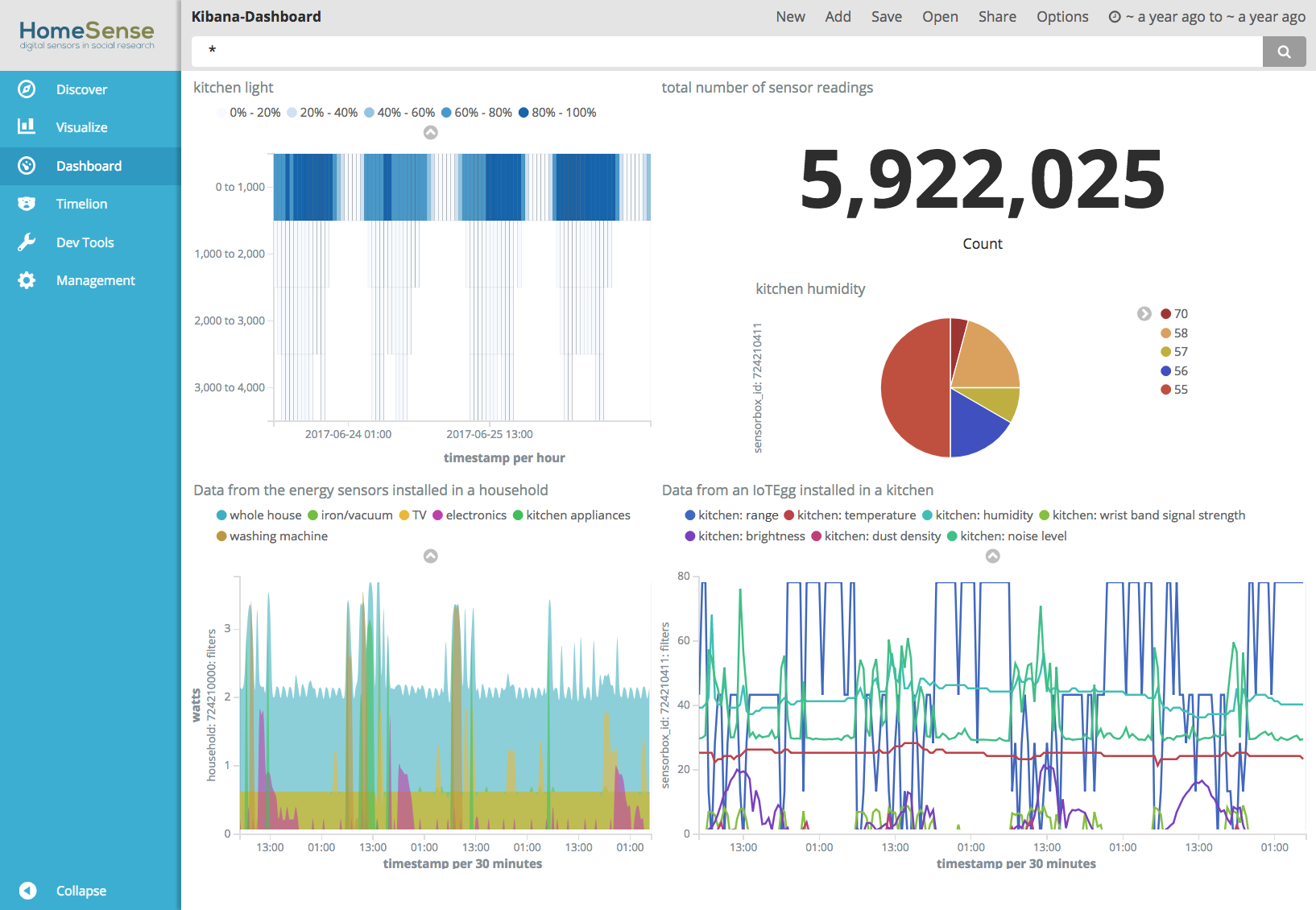}
  \caption{Kibana Dashboard.}
  \label{fig:kibana_dashboard}
\end{figure}

MakeSense relies on the open-source search engine ElasticSearch \cite{ElasticSearch:2018} and its dedicated analytics and visualisation platform Kibana \cite{Kibana:2018} for data visualisation and analysis. ElasticSearch supports scalable and real-time search with an HTTP web interface and schema-free JSON documents. Kibana provides visualisation capabilities on top of the large volumes of data indexed by an Elasticsearch cluster. Users can create different kinds of visualisations such as line, bar and scatter plots, pie charts and maps with the indexed data. Furthermore, users can combine different visualisations to create customised dashboards, as shown in Figure \ref{fig:kibana_dashboard}, which can be shared via web or integrated as HTML frames into existing web applications. An extended view of the lower two quadrants in Figure \ref{fig:kibana_dashboard} will be shown in Figure \ref{fig:hs-visualisation} and explained in the associated text.

As for data analysis, in addition to simple statistics (max, min, standard deviation, etc.), users may carry out more advanced data analysis exploiting Kibana's plugins such as X-Pack \cite{Xpack:2018} which enables the analysis of time-series data by creating models of normal behaviours in the data and identifying anomalous patterns in that data.

\section{Show cases}\label{sec:showcases}

\subsection{Home Monitoring}
A recent application of MakeSense was in the HomeSense project \cite{HomeSense:2018} which aimed at studying how digital sensors can be used by social researchers and facilitate their understanding of home life, i.e., what do people do at home, when, where, and alone or with others. For example, it is expected that when people cook hot meals, there are corresponding changes in signals such as temperature, humidity, electricity usage, etc. that can be captured by sensing devices. In the rest of the this section, we illustrate how MakeSense was used in the HomeSense project with respect to IoT device deployment and data analysis.    

\subsubsection{Installing Sensor Suites In Homes}
The HomeSense project recruited 20 households where sensor suites were installed to monitor the occupants' home activities. The suites of sensors used in HomeSense consist of the IoTEgg, the MiBand wristband and the CurrentCost energy monitor illustrated in Section \ref{sec:cotssensors}. 

\begin{figure}[!h]
\centering
  \includegraphics[width=0.9\textwidth]{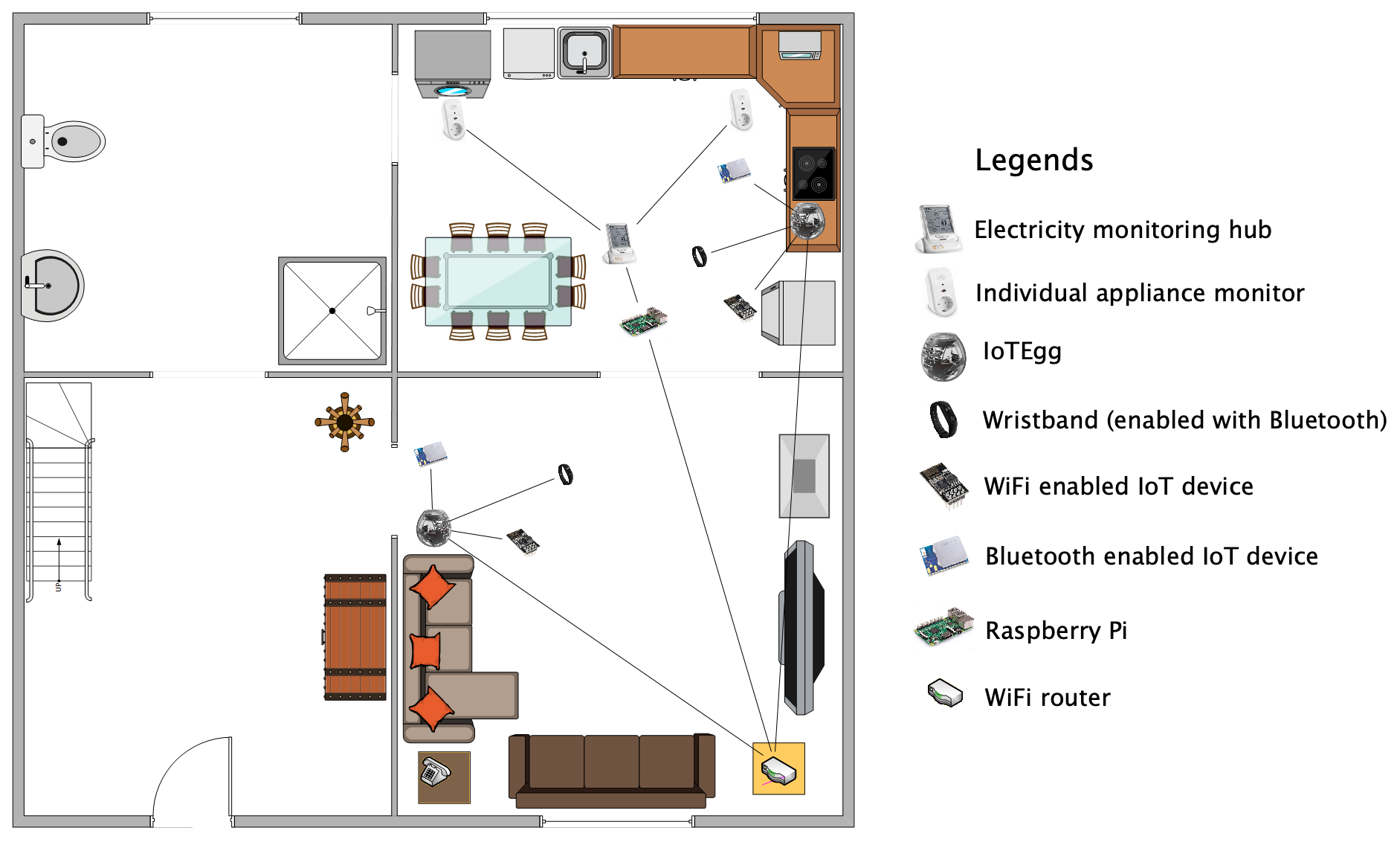}
  \caption{Example of Sensor Suites Installation at Home.}
  \label{fig:hs-install}
\end{figure}

With each household having a different configuration of physical spaces, household members, Internet connectivity, we had to plan the locations of deployments for the sensor suites such that the installation could be easily adjusted to different households. For example, Figure \ref{fig:hs-install} shows a floorplan of a household with spots marked for IoT device installation. The principle for installation is to cover as much as possible the space of the house such that the environmental changes caused by human activities can be captured. However, there are also cases where the participants asked explicitly to have certain rooms/places being excluded due to privacy concerns. The number of devices installed in each participating house ranged from 13 to 22 and their connectivity setup on average took about 10 minutes. MakeSense currently does not provide any mapping tool for drawing floor plans. For the HomeSense project, the researchers manually drafted the floor plan during the first visit to a participating household and marked down potential locations for installing sensors. The floor plans were then cleaned up, finalised and digitalised after the field deployment was completed. 

WiFi was available in most parts of each household, which makes it possible to transmit the data collected by the IoT devices in real time to the remote data server deployed on a cloud infrastructure managed by OpenStack. The data collected (every 3 seconds) from each IoTEgg are directly sent to the remote data server. The data collected (every 6 seconds) from the energy monitors are first directed to a Raspberry Pi via a USB to RS232/TTL PL2303HX cable and then sent to the remote data server while a local copy is also saved. Concerning the wristbands, household members were asked to wear them when at home so that all the IoTEggs installed in the house were able to detect the presence of each household member by means of the Bluetooth signal strength. Moreover, all the data are encrypted with AES (128 bits) and the household specific information is pseudonymised.

\subsubsection{Recognising Home Activities From Sensor-generated Data} 
\begin{figure}[!h]
\centering
  \includegraphics[width=\textwidth]{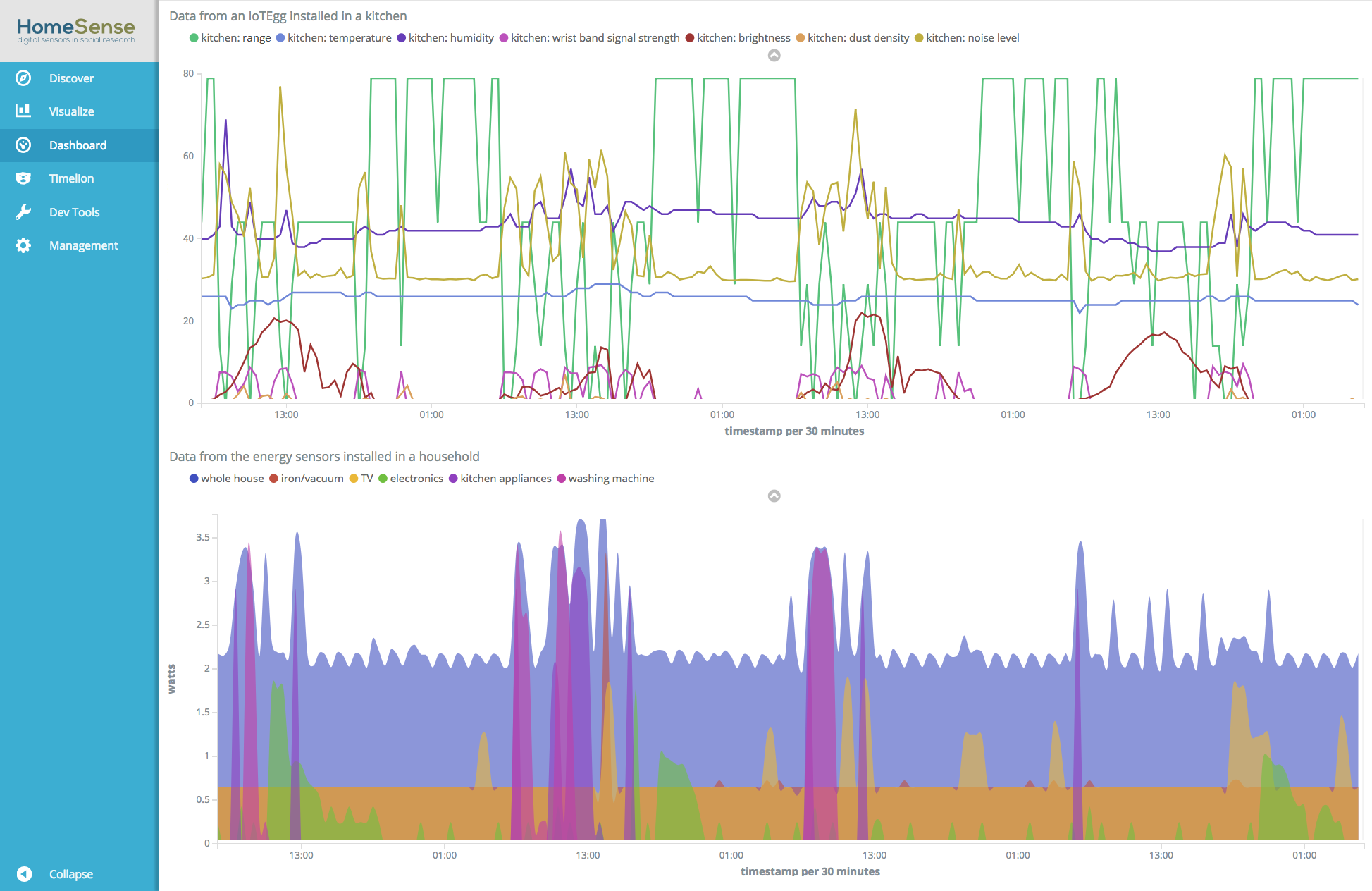}
  \caption{Visualisation of the data collected from the IoT devices (IoTEgg and Energy monitors).}
  \label{fig:hs-visualisation}
\end{figure}

In HomeSense, it was expected that the data collected from all the sensors deployed in different households could be used to help researchers better understand the activities occurring in the house and how they were assembled into household routines. To do so, it is essential that the researchers are capable of seeing the data in action and in a systematical way, i.e., how the data can be visualised is important. To assist social researchers easily visualise and analyse the sensor-generated data (in real time), MakeSense provides templates for visualising time series. For example, Figure \ref{fig:hs-visualisation} shows a screenshot of the sensor-generated data collected in one participating household via the visualisation module of MakeSense.

The visualisation of the data collected from the IoTEgg, as shown in the upper plot of Figure \ref{fig:hs-visualisation} shows the changes of signals in the kitchen of the household by measurements of range, temperature, humidity, wristband signal strength, brightness, dust density and noise level (indicated by line plots of different colors). We can see that the kitchen has been used several times during the day based on the observation that the ranging sensor detected moving objects in proximity (lower values indicating objects moving closer to the ranging sensor) and the signal strength of the wristband worn by the occupant has increased (higher signal strength indicating that the person wearing the wristband is closer to the IoTEgg). Accordingly, we can see that the noise level also increased. The changes in the humidity level may be caused by drink or food preparation. Notice that the ranging sensor outputs a maximum value of 80cm due to the narrow space in the kitchen and the location where the IoTEgg was placed. Moreover, the values of the wristband signal strength, brightness and dust density are all re-scaled using Kibana's scripted fields in order to integrate all the sensor information in one plot.  

The visualisation of the data collected from the energy monitors, as shown in the lower plot of Figure \ref{fig:hs-visualisation} indicates how much electricity the whole house has been using as well as the electricity consumption of the individual appliances including iron/vacuum, TV, electronics, kitchen appliances and washing machine via area plots of different colors. In relation to the visualisation of the data from the IoTEgg, we can see that the electricity usage of the kitchen appliances increased when there were movements detected by the IoTEgg in the kitchen. Moreover, there is also correlation between the electricity usage of the washing machine and the presence in the kitchen. This is due to the fact that the washing machine is located in the kitchen. 

\begin{figure}[!h]
\centering
  \includegraphics[width=0.7\textwidth]{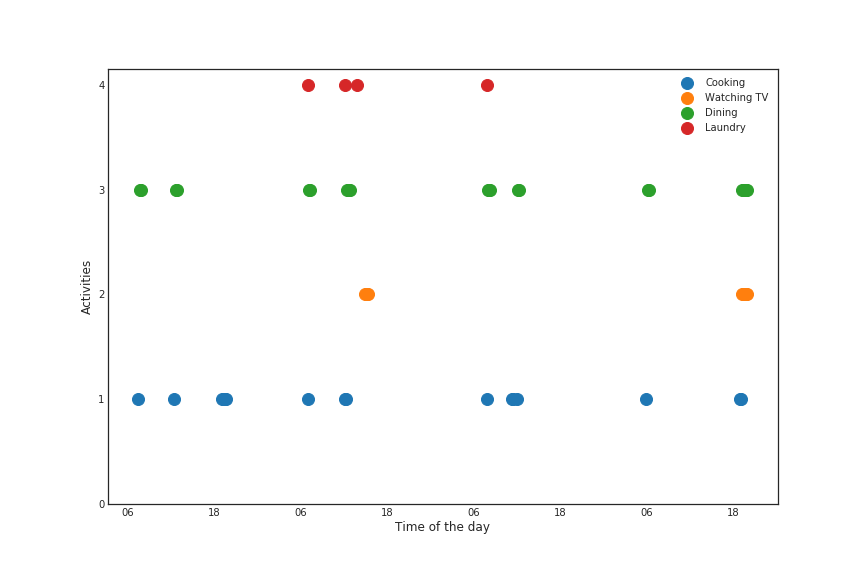}
  \caption{Visualisation of the time use diary.}
  \label{fig:hs-tud-visualisation}
\end{figure}

To evaluate whether our observations from the sensor-generated data are correlated to what happened in the house, the HomeSense project also collected time use diary from each household giving detailed information of what activities occurred where, when, by whom and with what types of appliances with an interval of every 10 minutes. Figure \ref{fig:hs-tud-visualisation} shows an example of the occurrences of four types of activity recorded in a time use diary including cooking, watching TV, dining and doing laundry, indicated by dots of different values and colors. For example, we can see that the occurrences of laundry activity (colored red with value 4) have a strong correlation with the electricity usage of the washing machine. 

Comparisons of the recordings obtained from the time use diaries and the knowledge extracted from the sensor-generated data not only help the researchers gain an in-depth understanding of the underlying environments where different activities were carried out but also help them validate their reasoning of whether certain activities occurred and/or how the activities occurred.  

With MakeSense, the social researchers were able to deploy a rich set of non-intrusive sensor suites in 20 real households. The data collected from the sensors provided the researchers with extensive information of the physical environments where the participants' daily routines are carried out. The visualisation of such data further helps the researchers understand the flow of household activities and validates their reasoning about the household practices, together with the information obtained from other methods such as time use diaries.   

\subsection{Office Monitoring}
Office environments can be monitored to ensure that the indoor environment quality and productivity standards are met \cite{AlHorr:2016}. Thermal comfort, ambient light levels, noise levels and indoor air quality and ventilation are some of the fundamental physical variables to measure to ensure the well-being of office occupants. 
In this section, we report on the work carried out with MakeSense in a real office setting, demonstrating how social scientists can benefit from the testbed to set up experiments to monitor and study human behaviour in an interactive way. 

\subsubsection{Deploying Sensor Suites In Offices}
\begin{figure}[!h]
\centering
  \includegraphics[width=\textwidth]{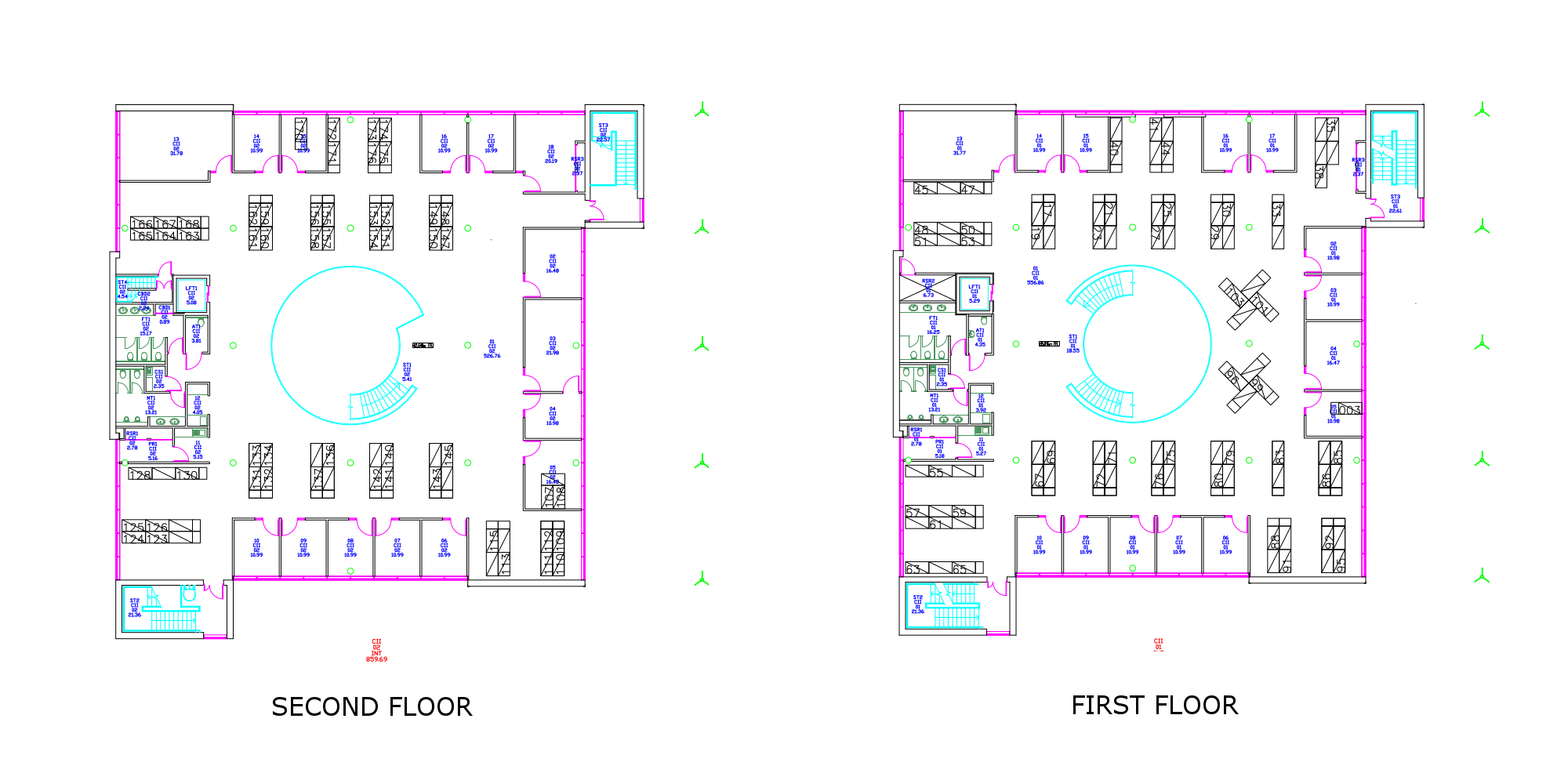}
  \caption{5GIC IoTEgg Desk Deployment.}
  \label{fig:5gic_depl}
\end{figure}

Figure \ref{fig:5gic_depl} shows the deployment of 100 IoTEggs in the 5G Innovation Centre (5GIC) at University of Surrey, with each IoTEgg placed on a numbered desk. The link between the IoTEgg LWM2M client ID and the desk number is randomised, thus making it difficult to trace the desk user.

Each numbered desk is assigned one IoTEgg which is placed on top of it, facing the user, and plugged into a nearby power socket and connected to the WiFi access points available in the 5GIC. The sensors measurements are collected every second from each IoTEgg by the M2M application running on a virtual machine (VM) deployed on the in-house OpenStack installation.

\subsubsection{Indoor Environment Quality Monitoring and Feedback}
The office monitoring system aims at studying how certain environmental variables influence the level of comfort and working conditions of the office occupants and provides an M2M application which is capable of providing feedback to users, e.g. via a buzzer actuator and an email notification service. This is achieved by leveraging the IoTEgg proximity sensor for estimating desk occupancy and the IoTEgg temperature, humidity, light and dust sensors for calculating the difference between the real-time sensor measurements and the user specified preferences. Moreover, users can configure his/her desk location (number) and a preferred email address for getting notifications about when the desired standards of comfort are violated, so that the user could take actions to make improvements.

\begin{figure}[!h]
\centering
  \includegraphics[width=\textwidth]{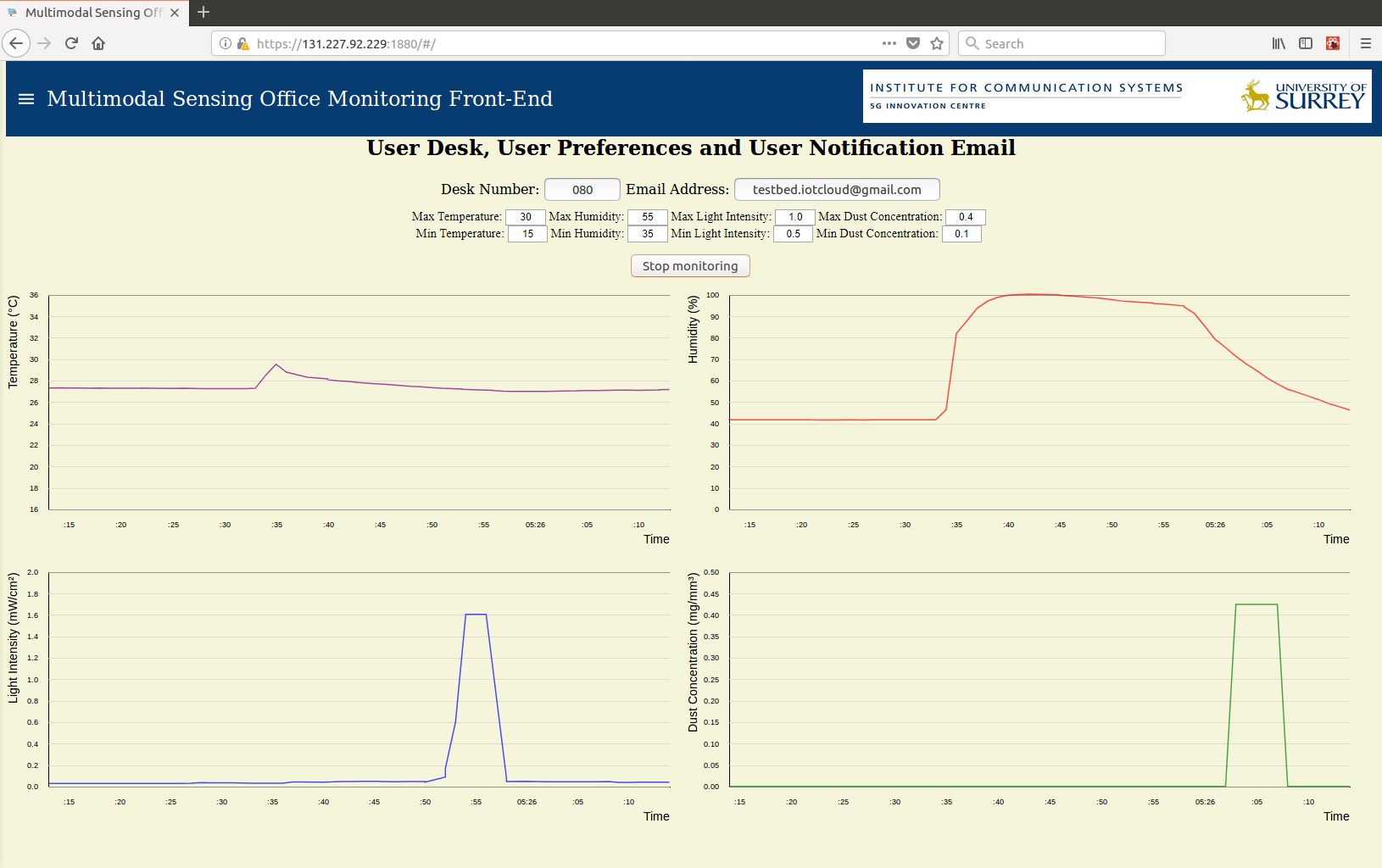}
  \caption{5GIC Multimodal Sensing Office Monitoring Live Data.}
  \label{fig:5gic_data}
\end{figure}

Figure \ref{fig:5gic_data} shows the interface via which users can specify their environmental preferences and monitor the changes of the environmental conditions. Six text boxes are provided for the users to configure their desk locations and desired values of environmental variables. A button allows to start/stop the monitoring for the configured desk. The peaks in the plots of the figure showcase the changes of the environmental conditions. 

When the monitoring starts, a background service verifies whether an user is in close proximity, i.e. by polling the proximity sensor for measurements less than 75cm which has been chosen empirically. If the user is in close proximity and the sensor values exceed the user preferences, notifications with customised messages are sent to the user and the IoTEgg buzzer is actuated once, suggesting the user to take certain actions to improve his/her working environment.
For example, suggestions for actions to be taken as a consequence of violating the desired working conditions can be provided as feedback emailed to the desk user, in order to improve the thermal comfort, light intensity level and air quality. Depending on the building type, suggestions could be opening the windows, adjusting the air conditioner or heating, asking for additional table lamps or operating the blinds when available.

In this use case, we use MakeSense to showcase how easy it is to write an M2M application and a Front-End in Node.JS and React.JS which benefits from the LWM2M APIs and the IoTEgg. Indeed, LWM2M allows to control data collection remotely and to write an application in the cloud for monitoring and control. Thus, by exploiting MakeSense, social researchers can both study and act upon people with feedback to them, by monitoring and designing M2M applications able to react to changing environmental conditions.

\section{Setup and Deployment}
In this section, we show the steps that a social researcher could follow to set up a home monitoring application using MakeSense with IoTEggs. These are:

\begin{enumerate}
	\item Step 1. Server Deployment
    \begin{itemize}
    	\item Time: < 1 hour
        \item Components: Data Collection Server, Data Storage Backends, Data Visualization Server
        \item Action: Running the script for deploying servers.
        \item Required Technical Skills: Linux command line (shell) operations.
        \item MakeSense Provides: Docker based server images, automated deploying script \footnote{https://github.com/jiejiang-jojo/make-sense}.
    \end{itemize}
	\item Step 2. IoTEgg Deployment
    \begin{enumerate}
    	\item Step 2.1 IoTEgg Configuration
        \begin{itemize}
    	    \item Time: < 5 minutes
            \item Action: Configuring IoTEgg ID, data server address, sampling rate, WiFi SSID and password; compiling and uploading firmware to IoTEgg.
            \item Required Technical Skills: Using the Mbed online Compiler or other text editors to edit the configuration file, compile and upload the firmware to IoTEgg.
            \item MakeSense Provides: Configuration file template.
        \end{itemize}    	
    	\item Step 2.2 Quality Test
        \begin{itemize}
    	    \item Time: $\sim$ 24 hours
            \item Action: Deploying a set of IoTEggs in a controlled environment and compare the data collected against each other.
            \item Required Technical Skills: Using Kibana to create indices and visualisations.
            \item MakeSense Provides: Kibana.
        \end{itemize}    
    	\item Step 2.3 Field Deployment
        \begin{itemize}
    	    \item Time: < 10 minutes
            \item Action: Connecting IoTEgg to WiFi and checking data collection.
            \item Required Technical Skills: Using Kibana to create indices and visualisations.
            \item MakeSense Provides: Kibana.
        \end{itemize}    	
    \end{enumerate}
\end{enumerate}

MakeSense aims at reducing the burden on social researchers of setting up the servers for receiving, storing and visualising data. It uses Docker technology \cite{Docker:2018} to pack all the components into docker images. This reduces the prerequisites of using MakeSense. Without Docker, researchers who want to use MakeSense have to install all the dependencies, such as Python, Flask and PostgreSQL, themselves and make sure they are correctly configured. Such effort may take a day or even longer depending on their experience. The Docker technology employed by MakeSense only requires to be installed and all the dependencies and configurations are pre-built into the images which can be deployed with a simple script within 30 minutes. This also reduces the amount of knowledge required for deployment to the simpler commands of the Linux Shell.

The setup of IoTEgg for deployment should follow a procedure for quality assurance. It is not uncommon to find some sensors or devices faulty. The individual sensors come with a quality guarantee and the assembling process also includes a quality test, but faults found in field deployment are very expensive and even more so in the data collecting phase. Thus it is recommended to have a quality test just before the field deployment. Such test will further reduce the possibility of faulty components entering experiments. The recommended way of conducting such tests is to deploy a set of IoTEggs in a controlled environment and compare the data collected from each IoTEgg in a 24 hour window.

There is no need for social researchers to understand how the source code works because the IoTEggs can be configured by editing a file in the source code to change for example, the sensor ID and server address fields. Such editing can be carried out with the online editor provided by MBed, a popular online IDE for IoT devices. With a few clicks the configuration and firmware can be written to the IoTEggs, which are then ready for field deployment. The sensor ID can carry more information than just a serial number. It is recommended to structure the sensor IDs in a way that can be pattern matched to help identify visualisations later in the experiments.

It is recommended to collect relevant information and prepare the IoTEggs before the deployment. An example is the WiFi configuration details needed for the IoTEggs to access the home Internet and send data back to the servers. Provided that the IoTEggs are pre-configured before the deployment, the field work will be as simple as finding a place for the IoTEggs, plugging them to power sockets and checking whether the data is coming into the servers via the data visualisation platform. This could be done in less than 10 minutes.    

\section{Discussion}
\label{sec:dis}
MakeSense was built with the aim of facilitating the use of IoT devices for indoor monitoring for social research. Based on its use in two showcase scenarios relating to home and office monitoring, we share some experience and discuss the limitations of MakeSense.

\subsection{Informed Consent}
For studies that involve human subjects, we have to ensure adequately informed consent from all types of participants including children and those who participate indirectly, e.g. visitors. The researchers should feel confident that the consent that they get is genuinely informed, i.e., participants actually understand what they are signing up for. We found that it is important to include demonstrations as part of the consent process to demystify the technologies.

\subsection{Minimum disturbances}
Even with informed consent, for social studies of indoor activities, especially in the setting of homes, it is essential that the deployment of IoT devices does not disturb people's daily routines. This is not only to reduce participation burden and dropout rate but also minimize the effect introduced by the IoT devices. Based on our field work experiences, there are several things to be noted: (1) non-intrusive IoT devices are preferred, (2) participants need control over the IoT devices, e.g., turning them off when there are visitors who do not want to be monitored, (3) the IoT devices should be easily integrated to the household infrastructure, e.g. power supply, internet connection, etc.    

\subsection{Data Security}
When it comes to the monitoring of people's daily lives, especially in the setting of real homes, people have a strong desire for data security. It is essential that the sensor data is securely transmitted and stored and all the information relating to the individuals is pseudonymised. Though the participants may be not familiar with the technical details about security, they often require the researchers to explain how their data is collected, saved and used. For example, in the HomeSense project, when demonstrating how various sensors work and what kinds of information they capture during the first visit to the participating house, some participants were interested to know how their data would be managed and used in the research. 

\subsection{Integrated Visualisation}
Social scientists often rely on methods such as questionnaire, survey, interview, time use diary to gather information about practices at home or work places. However, sensor-generated data is not always as straight-forward as the data collected from those methods. In order to understand the sensor data and use them for social science studies, it is essential that the social researchers can watch the data in action and observe the patterns visually. For example, the social researchers in the HomeSense project found it useful to visually explore the data from various sensor modules in Kibana and cross-validate their hypothesis with the activity recordings from the time use diaries.

\subsection{Design Trade-off}
Currently, MakeSense does not support recording audio or video information and mainly incorporates low-energy IoT devices (low data rate sensors). On the one hand, collecting, transmitting, storing and visualising audio/video data may pose technical challenges that need more sophisticated infrastructure support. On the other hand, capturing sensitive information such as audio/video in real-life settings such as homes or offices may raise serious privacy concerns and pose barriers for recruiting participants especially when there are children and guests involved. Therefore, both technical and ethical issues should be reviewed when sensors such as microphones and cameras are being considered. We are planning to extend MakeSense to cater for such needs in the future.

\subsection{Hardware and Software Testing}
Some COTS sensors can only function correctly in a limited scenario, which may not be obvious beforehand. 
For example, having the IoTEgg up against a wall or sitting on a table may influence the temperature and proximity sensors due to thermal effects and restricted location of power supply. The proximity sensors we use may also be severely affected by other sources of radiation (e.g. heat). It is important to identify the limitations of the sensors before deploying. Moreover, due to self-manufacturing with COTS components, there might be faulty parts in the hardware. Thus it requires a test procedure before deployment. A simple procedure can be used to put the devices in a reference environment and compare the generated data with some reference data.

\subsection{Software Development}
Durability test is also important along with functional testing, because the devices are expected to run without manual intervention for a long period of time and it would cost a lot of time and effort to repair or replace the device when it is malfunctioning. For example, in the beginning, some IoTEggs were found malfunctioning after days of operation without a clear indication about the software and hardware problems. In order to recover from and to debug such crashes, watchdogs and logging systems were implemented.

\subsection{Cloud Operation}
Concerning the cloud computing infrastructure, the platform is required to be highly available and reliable, so the design should be fail-safe, that is, countermeasures should be in place if services fail or lag behind. For example, we found out that logging and automatic re-spawning of services and/or message queue workers were very helpful in lowering manual intervention, improving reliability of data collection and helping to debug issues.

\subsection{Cost Efficient Design}
With the rapid development of IoT, a large number of electronic modules are available for researchers to build their own sensor suites. Besides, there are a great variety of commercial off-the-shelf sensing products in the market that provide plug-and-play sensing solutions. However, different electronic modules and commercial products often have a big difference in terms of cost. Even building one's own sensor suites, there are many choices of electronic modules for the same type of sensor with varying sensing capabilities such as accuracy, range, resolution, response time, etc. It is important to keep a balance between cost and quality. The development of the IoTEgg was to provide reasonable sensing capabilities in the scope of indoor environments and in the mean time keep the cost within \pounds50. Moreover, MakeSense tries to take advantage of open-source projects, which not only enables researchers to fine-tune the design but also reduces the overall cost.

\subsection{Automated Setup and Deployment}
Due to experience of previous works, a lot of time was devoted in designing measures to automate repetitive operations via scripting. This was useful because the IoTEggs needed to be deployed in large scale. For example, firmware over-the air was provided and can be triggered with one-liner script. This script builds the newly updated software and deploy it on the target IoTEggs over the WiFi and then reboot and self-program the IoTEggs, exploiting the external flash memory. 
Moreover, to reduce the prerequisites of using MakeSense, pre-built Docker images and scripts were provided to automate the server deployment process with simple commands of Linux Shell.

\section{Conclusions}
\label{sec:con}
In this paper we presented MakeSense, an IoT Testbed to facilitate social studies of indoor activities. Through two use cases, we showed that the testbed is non-intrusive, cost-efficient and easy to deploy, safekeeping the security and privacy of the participants via data encryption and pseudonymisation and capable of being adjusted to diverse experiment settings. 

MakeSense allows remote monitoring and control, supports extensible APIs for integrating COTS or self-developed sensor suites. The use of the IoTEggs benefit from widely available commodity sensors and industry standard WiFi/BLE radios, which provides flexibility and ease of setup in deployments. Flexible solutions for storage and visualisation enable real-time monitoring and high quality data collection.

Social research can benefit from the use of MakeSense by obtaining an additional layer of observation into people's activities in locations such as homes and offices. The data collected from the IoT devices can help the researchers understand the flow of indoor activities and validates their reasoning about the underlying practices, together with the information obtained from other methods such as time use diaries. With its cost-efficient and flexible design, MakeSense enables researchers to set up experiments at large scales and can be easily adapted to other indoor settings such as nursing homes, schools, restaurants, shopping malls, etc.        

For future work, we plan to extend MakeSense with more types of sensors and actuators such that the users can easily adjust the testbed for their needs. We also plan to use MakeSense for other experiments in different settings such as healthy ageing in which we will research on how IoT devices can be used to promote personalised interventions. 

\section*{Acknowledgement}

The authors thank Dr William Headley for the design and manufacture of the IoTEggs and thank Dr Kristr\'{u}n Gunnarsd\'{o}ttir for carrying out the field work of HomeSense and sharing the field work experience. The work was carried out as part of the ``HomeSense: digital sensors for social research'' project funded by the Economic and Social Research Council (grant ES/N011589/1) through the National Centre for Research Methods.

\bibliographystyle{unsrt}  
\bibliography{main}

\end{document}